\documentclass[pre,a4paper,showpacs,twocolumn,amsmath,amssymb,floatfix]{revtex4-1}

\usepackage{amssymb}
\usepackage{amsmath}
\usepackage{amsfonts}
\usepackage{bm}
\usepackage{latexsym}
\usepackage{graphicx,color}

\newcommand{\rsfig}[1]{
  \begin{center}
    \vspace{0.3cm}
    \includegraphics*[width=8.5cm]{{#1}}
  \end{center}
}
\newcommand{\rrsfig}[2]{
  \begin{center}
    \vspace{0.3cm}
    \includegraphics*[width=0.46\textwidth]{{#1}}
    \hfill{}
    \vspace{0.3cm}
    \includegraphics*[width=0.46\textwidth]{{#2}}
  \end{center}
}

\newcommand{\comment}[1]%
{\textsf{\textcolor{red}{#1}}}

\begin{document}



\title{Vibrational states and disorder in continuously compressed model glasses}

  
\author{F.~L\'eonforte}
\email[Email: ]{leonforte@theorie.physik.uni-goettingen.de}
\affiliation{Institut f\"ur Theoretische Physik, Georg-August-Universit\"at,
  Friedrich-Hund-Platz 1, 37077 G\"ottingen, Germany}


\begin{abstract}
We present in this paper a numerical study of the vibrational eigenvectors of a two-dimensional amorphous material, previously deeply studied from the point of view of mechanical properties and vibrational eigen-frequencies \cite{Wittmer02,Tanguy02,Leonforte05,Leonforte06}. Attention is paid here to the connection between the mechanical properties of this material in term of elastic heterogeneities (EH), and how these inherent heterogeneous structures affect the vibrational eigenvectors and their plane waves decomposition. The systems are analysed for different hydrostatic pressures, and using results from previous studies, a deeper understanding of the boson peak scenario is obtained. The vibrational spectrum of a continuously densified silica glass is also studied, from which it appears that the pulsation associated with the boson peak follows the same pressure dependence trend than that of transverse waves with pulsation associated with the EH characteristic size.
\end{abstract}

\pacs{
  61.43.Fs;
  63.50.Lm;
  62.30.+d.
}


\maketitle

\section{Introduction}
\label{subsec:summary}

Glasses develop a resistance to a macroscopic deformation, like solids, but depict a lack of structural order, like  liquids. In solids, the response to such a deformation is theoretically approached by the use of the continuum elasticity, which involve phonons to model the vibrational excitations. Due to the loss of translational invariance in glasses in comparison to the corresponding crystal, propagating phonons are difficult to describe, but have already been observed down to the atomic level \cite{Sette98}. It's also admitted that a common characteristic of disordered solids, when compared to crystalline ones, lies in the existence of an excess of low-frequency vibrational states, the well known boson peak. This excess produces the also well-known specific heat anomaly of glasses at temperatures $T\sim 10\,K$, and seems to be a signature of the disorder in glasses beyond the nanometer scale.

Many attempts have been proposed to theoretically interpret the boson peak anomaly \cite{Nakayama99,Oligschleger99,Vainer06,Grigera03,Lubchenko02,Maurer04,Schirmacher07,Gurevich03,Chumakov04}, and among them, Shintani and Tanaka \cite{Shintani08} have recently proposed an interpretation in term of involved atomic arrangements with transverse vibrations that tend to be particularly strong, in two-dimensional systems, around favoured structures of five-fold symmetry. The framework in which the boson peak is more related with transverse vibrations than longitudinal ones has been already approached in a series of papers dealing with the vibrational and mechanical properties of two- and three-dimensional soft (Lennard-Jones) and strong (silica) glasses \cite{Wittmer02,Tanguy02,Leonforte05,Leonforte06}. In those systems, transversal motions of the particles give rise to large vortices of characteristic size $\xi_{naff}$, when they respond to a macroscopic deformation. This noisy displacement field has been called \emph{non-affine} displacement field, and involve correlated displacement of atoms from $\xi_{naff}\sim 20$ inter-atomic distances in three dimensions, to $\xi_{naff}\sim 30$ in two dimensions. In both soft and strong glasses, the application of the classical continuum elasticity theory is subject to strong limitations below such length scales $\xi_{naff}$. This effect is particularly important when the wavelength of the vibrational excitation is less than this characteristic size $\xi_{naff}$. The origin of the departure from classical behaviour is very likely related to the disorder in inter-atomic interactions (local stresses, inhomogeneities in elastic constants, local anisotropy). This leads to an important decrease of the average shear modulus in amorphous systems compared with the corresponding crystals. Note that the typical size $\xi_{naff}$ depends on the pressure: it decreases and saturates at high pressures for both soft and strong glasses. Finally, it has been also previously noticed that the estimate of the frequency associated with $\xi_{naff}$ is in good agreement with the boson peak position \cite{Leonforte06}, and thus should encourage to consider the boson peak as a length marking a crossover between \emph{(i)} a regime where vibrations with wavelengths larger than $\xi_{naff}$ may be well described by a classical continuum theory, and \emph{(ii)} a small wavelength regime where vibrations are strongly affected by the elastic heterogeneities (EH).

In the following, such a picture will be confirmed in the case of a two-dimensional glass. After a summary of the theoretical concepts related with the continuum elasticity and an overview of the simulated system, the pressure dependence of the vibrational density of states (vDOS) and the boson peak will be approached for both strong and soft glasses. We then expect significant change of the vDOS upon compression, namely a decrease of boson peak height and its shift toward higher frequencies \cite{Deschamps09}. While this kind of pressure dependence study is merely used, in the literature, to improve the theoretical interpretations of the boson peak \cite{Geilenkeuser99,Sugai96,Hemley97,Inamura01,Monaco06,Niss07}, most of the present work will be devoted in the effect of increasing pressure, on the interplay between elastic heterogeneities (EH) and elastic wave propagation in a two-dimensional glass. 

\section{Technical details}

We first briefly review the considered systems studied in this work, much more details about them being given in previous works \cite{Wittmer02,Tanguy02,Leonforte05,Leonforte06}. Then, the theoretical framework used in background of this work is summarized.

\subsection{Simulated systems}
\label{subsec:systems}

The two-dimensional prototype materials we consider are bulk-like systems contained in a square of side $L$ with periodic boundary conditions. Computationally, they are formed by quenching very quickly a slightly polydisperse liquid of $N=10000$ spherical particles interacting via simple Lennard-Jones (LJ) pair potential, into the nearest energy minimum, following a fixed protocol~\cite{Wittmer02,Tanguy02} using standard molecular dynamics, steepest descent and conjugate gradient. The resulting structures are amorphous, {\it i.e.} they exhibit not a crystal-like, but a liquid-like order. The density of the bulk-like 2D glassy materials in the final state are adjusted in order to cover a wide range of hydrostatic pressure from $P=0\,\sigma^2/\epsilon$ to $P\sim 130\,\sigma^2/\epsilon$ (in LJ units). This kind of glass mimics a soft glass according to Angell's classification \cite{Angell95}. 
On the other hand, an amorphous silica glass is also simulated in the BKS framework \cite{BKS90}, and using a fixed quench protocol \cite{Leonforte06} that do not alter the final glassy state. For this kind of glass, systems containing between $N=24048$ and $N=64128$ particles were used, and for hydrostatic pressures from $P=0$ GPa to $P=20$ GPa.

\subsection{Vibrational modes from the continuous elasticity theory}
\label{subsec:vmodes}

In the framework of the linear classical theory of elasticity, which assume that elastic bodies are homogeneous and isotropic, the Landau free energy can be written on the form:

\begin{equation}\label{freeF}
	F=F_0 + \left(\frac{\lambda}{2} + \frac{\mu}{d}\right)\left(\mathrm{Tr}\underline{\underline{\epsilon}}\right)^2 + \mu\sum_{\alpha\beta}\left(\epsilon_{\alpha\beta} - \frac{\delta_{\alpha\beta}}{d}\mathrm{Tr}\underline{\underline{\epsilon}}\right)^2
\end{equation}

\noindent where $(\lambda,\mu)$ are the Lame coefficients, $d$ the space dimension, and the strain $\epsilon_{\alpha\beta}=\left(\partial_{\beta}u_{\alpha} + \partial_{\alpha}u_{\beta}\right)/2$ for the continuum displacement field $\underline{u}$. The stress is related to the strain using the classical relation $\sigma_{\alpha\beta}=\sigma^0_{\alpha\beta} + \lambda\mathrm{Tr}\underline{\underline{\epsilon}}\delta_{\alpha\beta} + 2\mu\epsilon_{\alpha\beta}$, where $\sigma^0_{\alpha\beta}$ is the reference state stress. Using this formalism, the equation of motion for the elastic field $\rho\partial_{tt}u_{\alpha}=\sum_{\beta=1}^d \partial_{\beta}\sigma_{\alpha\beta}(t) + f_{\alpha}$, can be written in a wave formulation:

\begin{equation}\label{wave1}
	\rho\frac{\partial^2 \underline{u}}{\partial t^2} = \left(\lambda + \mu\right)\nabla\left(\nabla.\underline{u}\right) + \mu\nabla^2\underline{u}
\end{equation}

\noindent The Eq.~\eqref{wave1} can be solved using the Stokes decomposition $\underline{u}=\nabla\phi + \nabla\times\psi$, with $\ddot{\phi}=C^2_L\nabla^2\phi$ for longitudinal waves of speed $C^2_L=\left(\lambda+2\mu\right)/\rho$, and $\ddot{\psi}=C^2_T\nabla^2\psi$ for transverse waves of speed $C^2_T=\mu/\rho$. These two wave equations can be solved using the plane waves basis with wave length $\lambda_p=2\pi/\parallel k\parallel$. For finite size systems as the ones we're simulating, $d$ quantum numbers appear and for instance in two-dimensions, $\left(k_x,k_y\right)=\left(2\pi/L\right)\left(n,m\right)$. Then, the vibrational density of states (vDOS) is expressed as:

\begin{equation}\label{DOS}
 	g(\omega)=\sum_s\int d\underline{k}(2\pi)^{-d}\delta(\omega - \omega_s(\underline{k}))
\end{equation}

\noindent Using the Debye's model \cite{Debye12} for set of phonon propagating in a continuum medium, this vibrational density of states can be explicitly evaluated, which leads to the famous formulae:

\begin{equation}\label{Debye}
	\quad g_D(\omega) = \left\{\begin{array}{ll}
	\text{2D:}\quad\frac{2\omega}{\omega^2_D}\text{,}\quad\omega_D = \left(\frac{8\pi\rho_{at}}{C^{-2}_T + C^{-2}_L}\right)^{1/2}\\
	\text{3D:}\quad\frac{3\omega^2}{\omega^3_D}\text{,}\quad\omega_D = 2\pi\left(\frac{9\rho_{at}}{4\pi\bigl(2.C^{-3}_T + C^{-3}_L\bigr)}\right)^{1/3}
	\end{array}\right.
\end{equation}

\noindent In particle based simulations, it's in principle possible to access the low temperature vibrational spectrum of glasses using the harmonic approximation $U_{Tot}=\sum_{i=1}^N E_{LJ}(r_{ij})\approx\sum_{i,j}M_{i\alpha,j\beta}u_{\alpha}(\underline{r}_i)u_{\beta}(\underline{r}_j)$. Hence, the equation of motion for an atom $i$, $m_i\frac{\partial^2 u_{\alpha}}{\partial t^2}(\underline{r}_i)=-\frac{\partial U_{Tot}}{\partial r_{i\alpha}}$, becomes equivalent for the whole system to an Euler-Lagrange eigenvalue problem $\left(m\omega^2 - \underline{\underline{M}}\right).\mathbb{Id}=0$ on the plane wave basis. Therefore, the vibration frequencies around an equilibrium position can be exactly calculated by diagonalizing the dynamical matrix $\underline{\underline{M}}$, which is expressible in terms of the first and second derivatives of the inter-particle interaction potentials, while the corresponding displacement fields are given by the eigenvectors.

\section{Results for the vibrational spectrum}
\label{sec:vib_spectrum}

Using the formalism developed in the previous Sec.\ref{subsec:vmodes}, a first look on vibrational eigen-frequencies is presented in the following, by also varying the applied hydrostatic pressure of simulated systems. The three-dimensional glass is approached using a silica glass, and the two-dimensional one using a Lennard-Jones glass. 

\subsection{Pressure dependence of the vibrational density of states}
\label{subsec:p_vDOS}

\begin{figure}
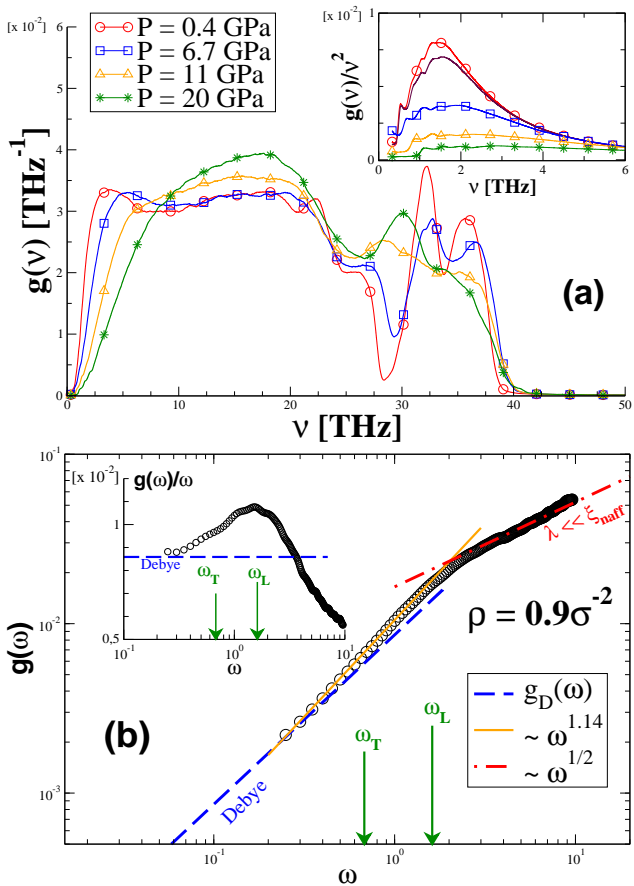

\rrsfig{fig1a.eps}{fig1b.eps}
\caption{Vibrational density of states (vDOS) for {\bf (a)} a silica glass at various pressures, and {\bf (b)} for a two-dimensional Lennard-Jones glass (LJ) at a density corresponding to a null hydrostatic pressure state. In the inset of both figures, the ratio to the Debye's prediction Eq.\eqref{Debye} is plotted, which shows the excess of vibrational density of state. On the {\bf (b)} part, the estimated contribution of longitudinal (L) and transverse (T) pulsations $\omega_{T/L}(P)\equiv 2\pi C_{T/L}(P)/\xi_{naff}(P)$ for the characteristic size $\xi_{naff}(P)$ \cite{Leonforte06} of the elastic heterogeneous domains are also drawn.}
\label{fig1}
\end{figure}

On Fig.\ref{fig1}(a), the vDOS from Eq.\eqref{DOS} for a silica glass under different hydrostatic pressure are plotted. Results are in agreement with numerical and experimental literature \cite{Pilla03,Niss07,Greaves07,Baldi09,Monaco06}. In the inset of this figure, the ratio to the Debye's prediction Eq.\eqref{Debye} which depicts the boson peak anomaly is shown. It appears that the boson peak intensity decreases with increasing pressure and tends to be unmeasurable at very high pressure. The corresponding boson peak frequency $\nu_{BP}$ also shifts to higher frequencies \cite{Deschamps09,Maurer04,Schirmacher07}.

\begin{figure}
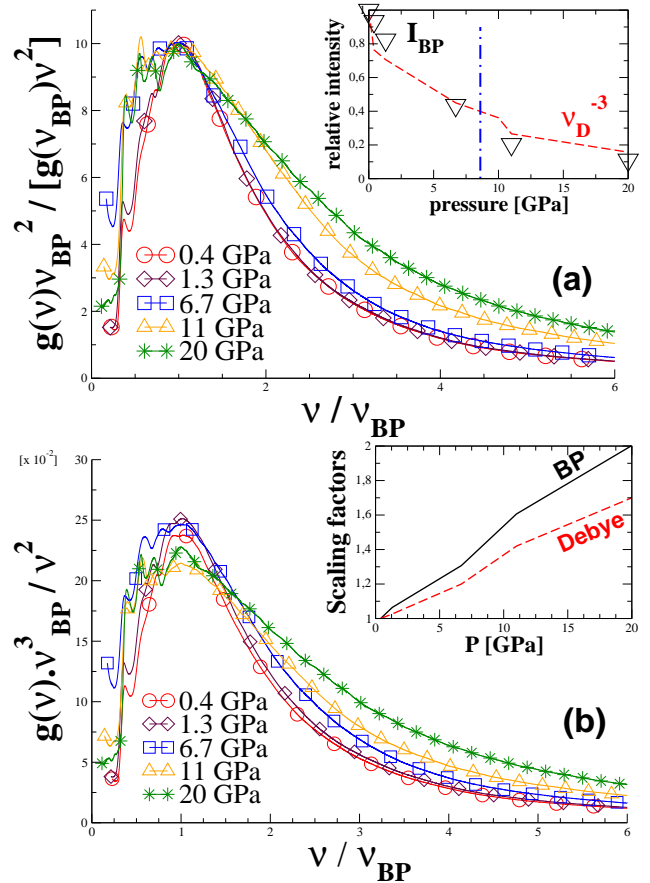

\rrsfig{fig2a.eps}{fig2b.eps}
\caption{Vibrational density of states for the silica glass and for various pressures, plotted in $\nu/\nu_{BP}(P)$ rescaled units, where $\nu_{BP}(P)$ is the boson peak frequency. Both scaling lead to a curve collapse for pressures less than $7$ GPa, above which the silica glass involves strong structural modifications. {\bf (a)}: vDOS are rescaled according to \cite{Niss07}. \emph{Inset:} relative intensity of boson peak and Debye level, normalized to unity by $P=0$ GPa values for each quantity. The dashed line marks the plastic transformation threshold upon compression \cite{Champagnon08}. {\bf (b)}: vDOS are rescaled to conserve modes \cite{Orsingher10}. \emph{Inset:} rescaled boson peak and Debye frequency (see text) versus pressure.}
\label{fig2}
\end{figure}

On the Fig.\ref{fig2}(a), a scaling analysis for the silica glass is plotted using reduced units $\nu/\nu_{BP}$, which may lead to a collapse of the vDOS on a single master curve \cite{Niss07}, at least depending the strength of the hydrostatic pressure. We then observe for pressures less than 7 GPa, that the rescaled vDOS collapse well (as also observed in \cite{Niss07}), but for higher pressures, this collapse is lost. This is related to a great change in the structure of the considered strong glass former at such pressures \cite{Trachenko04,Champagnon08}. In the inset of the Fig.\ref{fig2}(a), the variation of the boson peak intensity $I_{BP}(P)$ is compared to the one of the Debye level, while the onset of irreversible compression around $8.6$ GPa is also drawn \cite{Champagnon08}. In this inset, quantities are normalized to unity by their values at $P=0$ GPa. It has been observed experimentally \cite{Monaco06}, that the boson peak shift follows the variation of the Debye level under permanently densified silicate glasses, and so should be related to a transformation of the elastic medium. A contrary in \cite{Niss07}, the relative variation between both quantities has been shown to increase, i.e. the boson peak shift pressure dependence is stronger than for the Debye level. The authors in \cite{Niss07} thus conclude that the boson peak phenomena may not be only explained by a transformation of the elastic continuum upon compression.

\begin{figure}
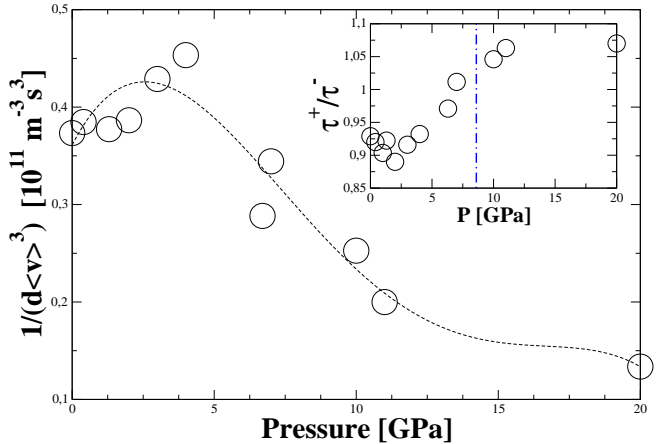

\rsfig{fig3.eps}
\caption{Variation of the continuous medium transformation (CMT) statement with hydrostatic pressure for a silica glass. A complex trend is observed, with a maximum close to the one measured in \cite{Deschamps09}. This points out that the CMT cannot solely explain the boson peak dependence with pressure. The dashed line is a guide for the eyes. \emph{Inset:} pressure variation of the ratio between identical rotations of inter-tetrahedral units and opposite ones, after an elastic elongation has been applied to a silica glass. At high pressure, identical rotations dominate, leading to a decrease of inhomogeneous particles rearrangements. Connected with the decrease of elastic heterogeneities characteristic sizes, this could explain the weakness of the boson beak intensity, while its shift could be approached by the one of speed of sounds for transverse waves.}
\label{fig3}
\end{figure}

In the inset of the Fig.\ref{fig2}(a), the relative ratio between $I_{BP}(P)$ and the Debye level decreases with the same trend upon compression, which should also corroborate the full continuum medium transformation (CMT) picture. In fact, the situation is more complex, as shown on the Fig.\ref{fig3}, where the CMT statement $g^{CMT}(\omega)/\omega^2\propto\left(d\langle v\rangle\right)^{-1}=(3d)^{-1}\left(C^{-3}_T+2C^{-3}_L\right)$ is plotted versus the compression level. As already pointed out in \cite{Deschamps09}, it's observed a strong non-linear pressure dependence of the CMT prescription in this silica glass. This suggests that the CMT picture cannot fully, or at least solely, explain the boson peak nature and behaviour under compression. A confirmation is then obtained on the Fig.\ref{fig2}(b), where the vDOS are plotted using another set of rescaling units $g(\nu).\nu^3_{BP}/\nu^2$ that takes into account the conservation of modes, while keeping the adimensional parameter $\nu/\nu_{BP}$ on abscissa. Again, reasonable collapse of vDOS is achieved for pressures less than $8.6$ GPa. More intriguingly, in the inset of Fig.\ref{fig2}(b) are also drawn the pressure variations of two scaling quantities with respect to the $P_0=0$ GPa state: $s^N(P)=\nu_{BP}(P)/\nu_{BP}(P_0)$ for the boson peak and $s^D(P)=\nu_{D}(P)/\nu_D(P_0)$ for the Debye frequency. As recently observed in \cite{Orsingher10} for a permanently densified $GeO_2$ glass, it appears that the rescaled frequency of the boson peak $s^N(P)$ has a slope greater than the Debye one $s^D(P)$, which also points out that the CMT cannot fully account for the pressure dependence of low-vibrational dynamics in silica glass.

Furthermore, it has been also shown in \cite{Deschamps09} by studying the half-width Raman intensity main band, that amorphous silica undergoes strong inter-tetrahedral structural rearrangements upon compression, which also shows that this glass is not homogeneous at nano-metric scale. We also measured the shift of the maximum of the inter-tetrahedra angle distributions with increasing pressure (not shown), and found values in agreement with experimental ones \cite{Deschamps09}.

In the inset of the Fig.\ref{fig3}, the ratio between the number $\tau_+$ of connected tetrahedral units that have the same rotation, and those $\tau_-$ that have opposite ones, is plotted versus the hydrostatic pressure, after an elastic elongation has been applied to the material. This ratio $\tau_+/\tau_-(P)$ is an indicator of the amount of inhomogeneous atoms rearrangements at a given hydrostatic pressure state, namely when $\tau_- > \tau_+$. In the inset of this figure, one can observes that the number of identical rotations $\tau_+$ has a tendency to dominate the number of opposite inter-tetrahedral moves, when the hydrostatic pressure increases. In that way, the silica glass undergoes a decrease of inhomogeneous particle rearrangements with increasing pressure, i.e. the response of the glass is more and more homogeneous at all scales. This could explain the decrease of the boson peak intensity, as well as also introduces the possible link between structural effects related to the disorder, and the vibrational properties of this glass.

Finally, on the Fig.\ref{fig1}(b), the vDOS of a Lennard-Jones glass under a null equivalent pressure density is also plotted, as well as the Debye prediction Eq.\eqref{Debye}. The difference that appears between the Debye's prediction and the computed vDOS is clearly drawn in the inset of the Fig.\ref{fig1}(b), where the ratio between both quantities depicts a bump around $\omega\sim 1\,rad.\tau^{-1}$, to which is associated the boson peak anomaly. Also drawn on this Fig.\ref{fig1}(b) are the estimated contribution of longitudinal (L) and transverse (T) pulsations $\omega_{T/L}(P)\equiv 2\pi C_{T/L}(P)/\xi_{naff}(P)$ for the characteristic size $\xi_{naff}(P)$ of the elastic heterogeneous domains, already discussed in the previous Sec.\ref{subsec:summary}. It then appears that the boson peak regime can be well approximated by the lower bound pulsation $\omega_T(P)$ of transverse waves associated with EH. 

\subsection{Pressure dependence of the boson peak pulsation}
\label{subsec:p_omegaBP}

\begin{figure}
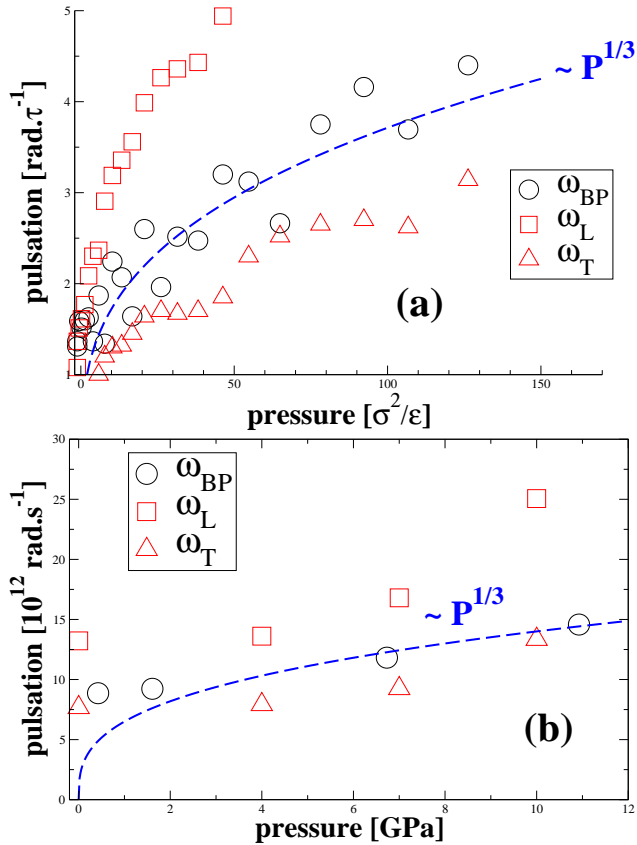

\rrsfig{fig4a.eps}{fig4b.eps}
\caption{Variation of the measured boson peak pulsation $\omega_{BP}(P)$ from Fig.~\ref{fig1} and for {\bf (a)} 2D Lennard-Jones and {\bf (b)} silica glasses, versus the hydrostatic pressure $P$. Boson peak pulsation is compared for each pressure to the estimation of the transverse $\omega_T(P)$ and longitudinal $\omega_L(P)$ pulsations associated with the elastic heterogeneities (EH) of characteristic size $\xi_{naff}(P)$. Note that from previous studies, the $\xi_{naff}(P)$ decrease with increasing pressure. The $P$-dependence prediction of Gurevich \emph{et al.} \cite{Gurevich03,Gurevich05} is also drawn, which quantitatively scales with $\omega_{BP}(P)$ and $\omega_T(P)$.}
\label{fig4}
\end{figure}

The last point discussed in the previous Sec.\ref{subsec:p_vDOS} is more precisely approached on the Fig.\ref{fig4}(a) and (b). On these figures, for both glasses the boson peak pulsation $\omega_{BP}(P)$ measured on the Fig.\ref{fig1} is plotted in comparison to the pulsations $\omega_{T/L}(P)$ for transverse and longitudinal elastic waves and for different pressures, i.e. using $C_{T/L}(P)$. Note that from previous studies, the characteristic sizes of EH for both glasses has been found to decrease and saturate at high pressure. On this figure, the power-law variation of the boson peak position $\omega_{BP}$ with $P$, as predicted by the soft potential model (SPM) \cite{Gurevich03,Gurevich05}, is also drawn. Our results seem to follow the expected $P^{1/3}$ variation from SPM, thus assuming that bulk modulus are pressure independent. For instance, in the case of silica, this situation is not fulfilled in the compression elastic regime. 

Then, on the same Fig.\ref{fig4}, we see that the transverse pulsations trend, with increasing pressure, follows the boson peak one, while discrepancies appear for longitudinal pulsations. This emphasizes previous results \cite{Tanguy02,Leonforte05,Leonforte06}: \emph{(i)} the response of a glass to a long wavelength excitation is dominantly transverse and leads to the creation of noisy rotational atomic rearrangements, \emph{(ii)} the characteristic size $\xi_{naff}$ of these structures decreases and saturates at high pressure, \emph{(iii)} the boson peak is sited close to the resonant regime of transverse vibrational modes interacting with these structures or in other worlds of pulsations associated with their characteristic size. This last point is mainly phenomenological, but it bridges the scale between the mechanical properties of glasses and their vibrational spectrum. In the next section, we'll go deeper in this understanding by bridging the scale to the vibrational eigenvectors.

\section{Results for the vibrational eigenvectors in 2D}
\label{sec:vib_ev}

As discussed in Sec.\ref{subsec:vmodes}, the vibrational eigenvectors in the framework of the continuous elasticity theory are plane waves. In a disordered glassy material, one can expect plane waves in the highest wavelength limit, while things may be more complicated with the decrease of the wavelength. Nevertheless, recent progress in computer simulations and experimental techniques have provided new insights in the understanding of this problem. For instance, in silica glasses \cite{Ruffle06,Courtens03,Ruffle03}, it has been shown that acoustic phonon waves undergo a crossover between a propagation regime and another in which they are strongly diffracted. This crossover corresponds to the Ioffe-Regel limit where the phonons mean free path is of same order than their wavelength \cite{Ioffe60,Ruffle06}, while the boson peak has been found to be closely related to this crossover frequency. It then appears that the boson peak seems to be more or less located in a region close to the Ioffe-Regel crossover \cite{Monaco09,Taraskin02,Maurer04,Chumakov04}, with wave excitations of predominantly transverse character. Recently, numerical studies \cite{Shintani08,Monaco09} have been employed to confirm this last point, by computing the dynamic longitudinal and transverse structure factors (only the former is accessible experimentally in the studied frequency range) from the knowledge of particle trajectories. In \cite{Shintani08}, the authors also showed using per particle vDOS, that transverse modes involved in the boson peak are associated with defective structures in the materials. 

In order to connect previous works related with EH in two-dimensional Lennard-Jones glasses (LJ2D) \cite{Tanguy02,Leonforte05,Leonforte06} and their characteristic size $\xi_{naff}(P)$, the problematic is here approached by the use of the eigenvectors and their interaction with such specific zones.

\begin{figure}
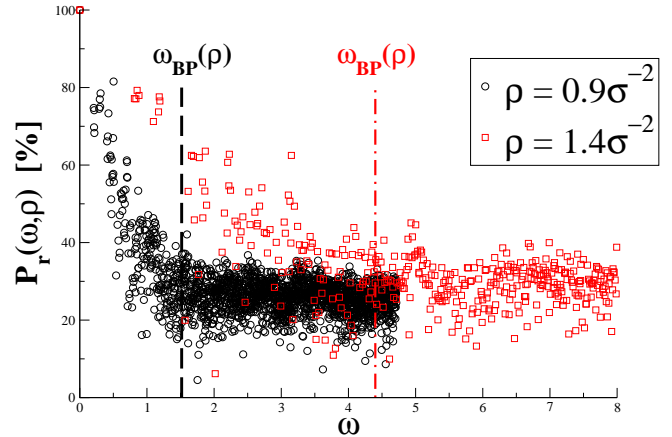

\rsfig{fig5.eps}
\caption{Participation ratio $P_r(\omega,\rho)$ from Eq.\eqref{Prp}, for two LJ2D systems under different densities. The corresponding pressure states are $\rho=0.9\,\sigma^{-2}\Leftrightarrow P\sim 0\,\sigma^{2}/\epsilon$ and $\rho=1.4\,\sigma^{-2}\Leftrightarrow P\sim 130\,\sigma^2/\epsilon$. Only the first $2000$ eigenvectors were used for this plot, and the boson peak pulsation $\omega_{BP}(\rho)$ from Fig.\ref{fig1}(b) is also plotted. The change in the participation regime trend is marked by the boson peak pulsation, measured on the vDOS. 
}
\label{fig5}
\end{figure}

To this aim, we first analyse the participation ratio of eigenvectors along the vibrational spectrum. This one is obtained by computing the quantity:

\begin{equation}\label{Prp}
	P_r(\omega,\rho) = N^{-1}\frac{\sum_i^N \mid \mathbf{e}_i(\omega,\rho)\mid^2}{\sum_i^N \mid\mathbf{e}_i(\omega,\rho)\mid^4}
\end{equation}

\noindent where $\mathbf{e}(\omega,\rho)$ are the eigenvectors for a pulsation mode $\omega$ under an overall hydrostatic pressure $P$ or density $\rho$. When a mode is delocalized and then most of atoms vibrate to the corresponding pulsation $\omega$, $P_r(\omega,\rho)\approx 1$, however, if the mode is localized, then few atoms participate to the mode and $P_r(\omega,\rho)\rightarrow N^{-1}$. The participation ratio is plotted on the Fig.\ref{fig5} for two density values corresponding to a free $\rho=0.9\,\sigma^{-2}\Leftrightarrow P\sim 0\,\sigma^2/\epsilon$ state and a strongly compressed one with $\rho=1.4\,\sigma^{-2}\Leftrightarrow P\sim 130\,\sigma^2/\epsilon$. On this figure, it appears that $P_r(\omega,\rho)$ decreases from a full atomic vibrational participative state to a lower one that is stabilized around $30\%$ over the studied pulsation range. Let us remind that for the $P\sim 0\,\sigma^2/\epsilon$ sample, the Debye pulsation is around $\omega_D(P)\approx 15.3\,rad.\tau^{-1}$. Then, one may expect for higher pulsations $P_r(\omega,\rho)$ to tend to $N^{-1}$. A crossover appear in $P_r$ at a pulsation close to the boson peak one (measured using the vDOS) which means that around the boson peak, a clearly non-null amount of atoms participate to the vibrational states. 

\begin{figure}
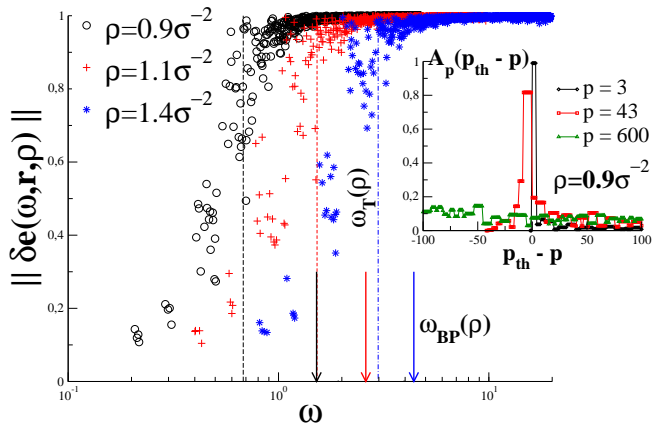

\rsfig{fig6.eps}
\caption{Noise in eigenvectors from Eq.\eqref{dvec} versus pulsation $\omega$ for different glass densities in 2D. For each density, the measured boson peak pulsation $\omega_{BP}(\rho)$ is marked with a vertical arrow, while the vertical dashed line is the value of the pulsation $\omega_T(\rho)=2\pi C_T(\rho)/\xi_{naff}(\rho)$ that would have a transverse wave of wavelength of the same size than $\xi_{naff}(\rho)$. The noise becomes predominant for pulsations close to $\omega_T(\rho)$ and saturates when the boson peak $\omega_{BP}(\rho)$ is reached. \emph{Inset:} projected amplitudes over plane waves from Eq.\eqref{Apq}, and for a null equivalent hydrostatic pressure. The distance to the theoretical plane wave is given by $(p_{th}-p)$: $p=3$ is the first mode after the two Goldstone's ones, $p=43$ is somewhere before the boson peak, and $p=600$ is far after the boson peak. Spreading of the projected amplitude over neighbouring modes is observed, which increases with the considered mode index $p$. For the highest mode $p=600$, the plane wave decomposition does not hold.}
\label{fig6}
\end{figure}

A closer look to the eigenvectors can be obtained by defining a pseudo-spectral density for an eigenstate $\mid \mathbf{e}(\omega_p,r,\rho) \rangle$ for a mode $p$ of pulsation $\omega_p$, over a {\it theoretical} plane wave of wave-vector $\mathbf{k}_q$, given for a mode $q$ of pulsation $\omega_q$:

\begin{equation}\label{Apq}
	A_p(q) = \langle \mathbf{e}(\omega_p,r,\rho) \mid \mathbf{e}_{theo}(\omega_q,r,\rho)\rangle
\end{equation}

\noindent With the above definition, one can build the noisy eigenvector $\delta\mathbf{e}(\omega_p,r,\rho)$:

\begin{equation}\label{dvec}
	\delta\mathbf{e}(\omega_p,r,\rho) = \mathbf{e}(\omega_p,r,\rho) - \sum_{q\in\mathcal{D}_p} A_p(q)\mathbf{e}_{theo}(\omega_q,r,\rho)
\end{equation}

\noindent where $\mathcal{D}_p$ is the degeneracy set for the mode $p$ of pulsation $\omega_p$. On inset of the Fig.\ref{fig6}, the pseudo-spectral density Eq.\eqref{Apq} is plotted for the $P\sim 0\,\sigma^2/\epsilon$ sample, versus the "distance" $(p_{th}-p)$ to the theoretical mode $q=p_{th}$, and for three modes: $p=3$ is the first after the two first Goldstone's modes, $p=43$ is an arbitrary mode before the boson peak, and $p=600$ is far away after the boson peak mode. In the inset of this Fig.\ref{fig6}, one can note that the lowest mode amplitude is well stuck over the theoretical plane wave, while increasing the mode index leads to a spreading of the projected amplitudes $A_p(p_{th}-p)$. This means that the plane wave picture used to describe eigenvectors is more and more lost when the mode index increases, or at least that as soon as the mode index increases, more and more neighbouring waves pollute the ideal theoretical plane wave which should represent the $p$ mode. This can be more clearly seen by computing the noise in eigenvectors from Eq.\eqref{dvec}. Results are shown on the main panel of the Fig.\ref{fig6} for three different hydrostatic pressures. The trend of $\delta\mathbf{e}(\omega_p,r,\rho)$ is compared for each density $\rho$ to the boson peak pulsation $\omega_{BP}(\rho)$, and to the pulsation $\omega_T(\rho)=2\pi C_T(\rho)/\xi_{naff}(\rho)$ of resonant transverse waves of wavelength equal to the EH characteristic size $\xi_{naff}(\rho)$. It then appears that the noise becomes predominant for pulsations around $\omega_T(\rho)$ and that starting from the boson peak regime, the plane wave picture no more holds \cite{Courtens03,Ruffle03,Taraskin02,Maurer04}.

\begin{figure}
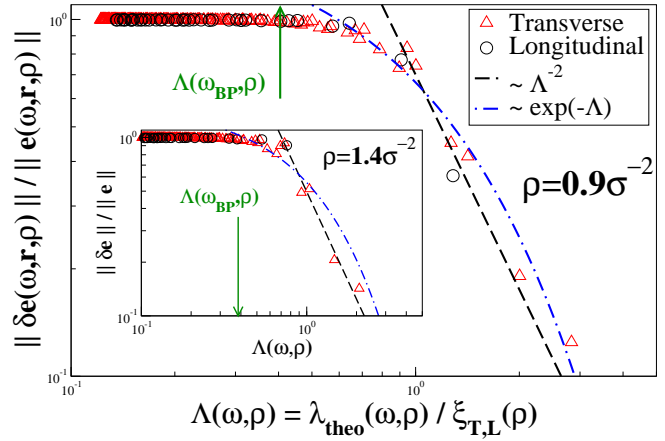

\rsfig{fig7.eps}
\caption{Relative noise in eigenvectors as a function of the reduce wavelength $\Lambda(\omega,\rho)$ (see text for details), and for two samples under different density. The Rayleigh regime $(\sim\Lambda^{-2})$, where plane waves are diffracted by the weak disorder, holds for wavelengths greater than EH size. An intermediate exponential regime forms a crossover for wavelengths between EH size and boson peak wavelength, while strong diffraction occur for lower wavelengths.}
\label{fig7}
\end{figure}

An interesting point also results from the quantity $\omega_T(\rho)$. If for all considered densities, this one lies just before the noisy saturated regime, one can also wonder how the EH affect the plane wave framework and thus are involved in the propagation of elastic waves. On the Fig.\ref{fig7}, the problematic of the interaction process of plane waves with EH is then approached by studying the noise relative ratio $\parallel\delta\mathbf{e}(\omega,r,\rho)\parallel/\parallel\mathbf{e}(\omega,r,\rho)\parallel$. On this figure, a reduced wavelength $\Lambda(\omega,\rho)=\lambda_{theo}(\omega,\rho)/\xi_{L,T}(\rho)$ is used, with $\xi_{L,T}(\rho)$ the scaling lengths from the 2D scaling mode analysis \cite{Tanguy02} for longitudinal (L) and transverse (T) waves, and $\lambda_{theo}(\omega,\rho)$ the theoretical wavelength for a plane wave at a given pulsation $\omega$ in the continuum medium under a density $\rho$. Note that we have observed that the EH characteristic size $\xi_{naff}(\rho)\simeq\xi_T(\rho)$ while $\xi_L(\rho)\sim 2\xi_T(\rho)$, and that $\omega^{\star}_L(\rho)\simeq\omega^{\star}_T(\rho)$ with $\omega^{\star}_{L,T}(\rho)=2\pi C_{L,T}(\rho)/\xi_{L,T}(\rho)$. Hence, with such a choice of units, one can remove the longitudinal and transverse wave types under an unique plot, while keeping in mind that one can recover true units by using the relation between $\xi_{naff}(\rho)$, and $\xi_L(\rho)\sim 2\xi_T(\rho)$. On the same figure, the reduce wavelength for the boson peak $\Lambda(\omega_{BP},\rho)$ is also drawn, with the true wavelength $\lambda(\omega_{BP},\rho)=2\pi C_{T,L}/\xi_{T,L}(\rho)$. Again, using the reduced wavelength $\Lambda(\omega,\rho)$, the boson peak wavelength thereby estimated is the same for $(L,T)$ waves. On this Fig.\ref{fig7}, it appears three regimes for the relative noise amplitude: \emph{(i)} for wavelengths greater than the EH characteristic size, a Rayleigh type diffusion process is observed, i.e. planes waves are scattered by a weak disorder, \emph{(ii)} for wavelengths between the boson peak one and ones of the same size than EH, an intermediate exponential regime seems to occur, \emph{(iii)} for wavelengths lower than the boson peak one, the noise is of same order than the eigenvector and plane waves are strongly scattered by the disorder. The same scenario holds for an higher pressure sample, as shown in the inset of the Fig.\ref{fig7}.

A useful way to characterize more precisely the interaction process between plane waves and disorder, is to define a scalar correlation function of the noisy eigenvectors given in Eq.\eqref{dvec}. This one is then written as:

\begin{equation}\label{corrdvec}
	C_{\omega}(\mathbf{r},\rho)=\langle\delta\mathbf{e}(\omega,r,\rho)\mid\delta\mathbf{e}(\omega,0,\rho)\rangle
\end{equation}

This correlation function depicts an anti-correlation zone at a distance $r=\langle\zeta^1(\rho)\rangle_{\omega}$ where the average is taken over the degeneracy set $\mathcal{D}_p$ for modes $p$ of same pulsation $\omega_p\equiv\omega$. This crossover from a positive correlation to a negative one, which thus define the first zero $\langle\zeta^1(\rho)\rangle_{\omega}$, is a track of the emergence of a dominant rotational structure in the noisy eigenvectors $\delta\mathbf{e}(\omega,r,\rho)$, and $\langle\zeta^1(\rho)\rangle_{\omega}$ can then be considered as a characteristic or threshold mean free path length for phonons propagation.

\begin{figure}
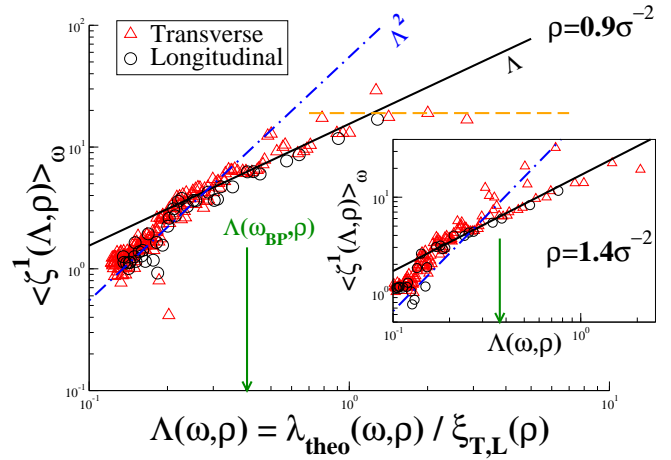

\rsfig{fig8.eps}
\caption{Dominant rotational structure size in noisy eigenvectors $\delta\mathbf{e}(\omega,r,\rho)$ from the first zero of the scalar correlation function Eq.\eqref{corrdvec}, versus the reduced wavelength $\Lambda(\omega,\rho)$ for T and L waves (remind that $\xi_L\sim 2\xi_T$). $\langle\zeta^1(\rho)\rangle_{\omega}$ is equivalent to the threshold mean free path length for propagating phonons in disordered media. Three regimes occur with crossovers between each marked by, first EH characteristic size at $\Lambda(\omega,\rho)=1$, second the boson peak position $\Lambda(\omega_{BP},\rho)$. The boson peak is sited at the Ioffe-Regel \cite{Ioffe60} limit. \emph{Inset:} same plot for system under an equivalent pressure of $P\sim 130\sigma/\epsilon^2$.}
\label{fig8}
\end{figure}

On the Fig.\ref{fig8}, the length $\langle\zeta^1(\rho)\rangle_{\omega}$ is plotted as a function of the reduced wavelength $\Lambda(\omega,\rho)$ for both longitudinal (L) and transverse (T) waves, and for two different sample density. On the same plot, the boson peak position $\Lambda(\omega_{BP},\rho)$ is drawn. Again, three regimes are observed: \emph{(i)} for phonon wavelengths greater than the characteristic length of EH at $\Lambda(\omega,\rho)=1$, the threshold mean free path length is independent of the mode, meaning that plane waves propagation is not strongly altered by the disorder; \emph{(ii)} $\Lambda(\omega,\rho)=1$ marks an independent/dependent crossover until $\Lambda\rightarrow\Lambda(\omega_{BP},\rho)$, and during which a linear dependence of the threshold mean free path length $\langle\zeta^1(\rho)\rangle_{\omega}$ with the wavelength appears: this is equivalent to the Ioffe-Regel regime \cite{Shintani08,Monaco09,Ioffe60}; \emph{(iii)} for wavelengths lower than the boson peak one, the strong diffraction regime of plane waves by the disorder is reached, with a quadratic dependence of the mean free path length with phonon wavelength. The boson peak then marks a crossover at the Ioffe-Regel limit \cite{Shintani08,Courtens03,Ruffle03,Taraskin02,Maurer04}. Note that the same scenario also holds for a more compressed sample as shown in the inset of the Fig.\ref{fig8}.

\begin{figure}
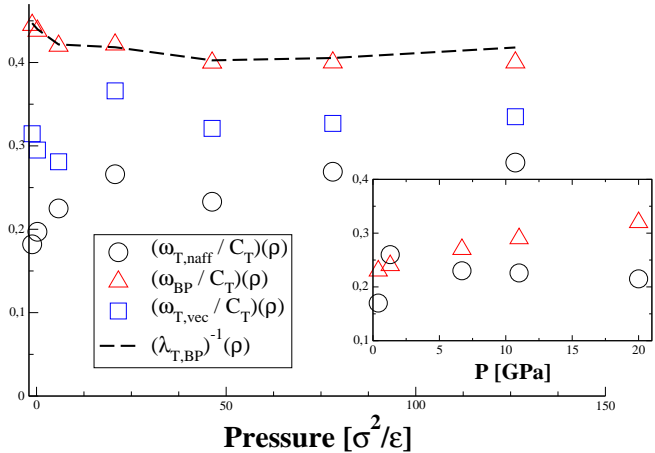

\rsfig{fig9.eps}
\caption{Plot of the inverse lengths from $(\omega_X/C_T)(\rho)$ ratio. Depending $X$ observable, lengths represent: EH characteristic size with $\omega_{T,naff}$, boson peak pulsation $\omega_{BP}(\rho)$, lowest mode threshold mean free path for elastic waves with $\omega_{T,vec}$. \emph{Main panel:} for the LJ glass, comparison is also shown with the boson peak wavelength, independently computed at $\omega=\omega_{BP}$ from the transverse eigenvectors power spectrum given in Eq.\eqref{ptspectrum}. The weak variation with varying pressure points out the fact that the boson peak is a signature of a characteristic length which emerges from an ordering/frustration competition in glasses. \emph{Inset:} same plot for the silica glass.}
\label{fig9}
\end{figure}

Finally, on the Fig.\ref{fig9}, results are summarized for different quantities encountered during this work. On this figure, data are plotted as the inverse of a specific length, namely by considering the ratio $(\omega_X/C_T)(\rho)\propto l^{-1}(\rho)$. As previously discussed in \cite{Leonforte06,Shintani08,Monaco09}, and in agreement with the present work, it appears that transverse modes are more affected by the disorder, as well as more involved in the boson peak anomaly than longitudinal ones. Consequently, this motivates the choice of transverse speed of sound $C_T(\rho)$ to perform the above ratio. The quantities $X$ are related to: {\bf (a)} EH characteristic size and $X:=(T,naff)$, {\bf (b)} boson peak and $X:=(BP)$, {\bf (c)} the threshold mean free path $\langle\zeta^1(\rho)\rangle_{\omega}$ in the mode independent regime, i.e. for wavelengths larger than $\xi_{naff}(\rho)$, and $X:=(T,vec)$. On the Fig.\ref{fig9}, we also plotted the inverse boson peak wavelength computed from the transverse power spectrum:

\begin{equation}\label{ptspectrum}
	f_T(\omega,\mathbf{k},\rho)=N^{-1}\biggl\langle\Bigl\lvert\sum_{j=1}^N \mathbf{\hat{k}}\wedge\mathbf{e}(\omega,\mathbf{r}_j,	\rho)e^{i\mathbf{k}.\mathbf{r}_j}\Bigr\rvert^2\biggr\rangle
\end{equation}

\noindent where $\mathbf{\hat{k}}=\mathbf{k}/\parallel\mathbf{k}\parallel$. Then the transverse boson peak wavelength is computed by $\mathrm{MAX}_{\mathbf{k}}\lbrace f_T(\omega_{BP},\mathbf{k},\rho)\rbrace$, giving $\parallel\mathbf{k}_{BP}\parallel=2\pi/\lambda_{T,BP}(\rho)$. Hence, on the Fig.\ref{fig9}, the pressure dependence of all these quantities are drawn. Interestingly, the inverse boson peak length $\lambda_{T,BP}(\rho)$ extracted from Eq.\eqref{ptspectrum} seems to not be altered by the pressure variation, and is very close to the inverse boson peak length computed using transverse speeds of elastic waves. The comparison between both independent measurements seems to point out the fact that the pulsation shift of the boson peak is directly related to the increase of speed of elastic waves under compression. We then also point out that the same trend seems to occur for inverse lengths computed from EH characteristic size and lowest mode threshold mean free path for elastic waves. In the inset of the Fig.\ref{fig9}, the same plot is drawn for the silica glass, and the same trend can also be observed. This type of plot should then encourage to view the boson peak anomaly as a process that results from the existence of a characteristic length, itself inherent from the competition between ordering and frustration in glassy materials.

\section{Concluding remarks}

During this work, we have extended to the pressure dependent regime, previous studies on a prototypical two-dimensional soft glass former. Taking advantage of known results concerning its mechanical response to a macroscopic excitation, and the relation it exists between the involved specific \emph{non-affine} atomic displacements (which are connected with the existence of EH) and its vibrational spectrum, we have developed an additional point of view on the boson peak scenario. This one has been achieved by studying the interplay between the vibrational eigenvectors and the well characterized EH, for this kind of system under continuous compression. 

We have demonstrated that eigenvectors with wavelengths larger than the EH characteristic size can be well approximated by plane waves, as predicted by the continuum theory of elasticity. Conversely, when the wavelength becomes of same order than this characteristic size, a crossover occurs and the plane wave framework no more holds. Hence, elastic waves are strongly scattered by the EH, while this crossover is marked by the boson peak pulsation, that also coincides with the Ioffe-Regel limit. 

We have also shown that the boson peak pulsation shifts with the pressure on a same way than the pulsation of transverse elastic waves of wavelength of same size than EH, for both soft LJ and strong silica glasses. This results confirms the fact that, whatever the pressure, the boson peak is always sited close to the \emph{non-affine} regime, where the response of the system to a large wavelength excitation is self-organized in a collective manner and in such a way that it emerges a characteristic length as the result of the competition between order and frustration in glasses. Then, we also pointed out that the pulsation shift of the boson peak is directly related to the increase of the speed of sound for transverse elastic waves. It follows that the boson peak can be indeed viewed as the result of a length, which might be related to this emerging length.

A part of this work was also devoted on the preliminary study of the pressure dependence of the eigen-frequencies of a silica glass, a strong glass former. The density of state of such a glass was computed for various pressures, while the mechanical response of this glass to a macroscopic elongation was also performed in a parallel independent study, in order to extract the characteristic size of EH. We have then measured a decrease and a saturation of this characteristic length with increasing pressure. By the same way, it has been also possible to extract the inter-tetrahedral reciprocal rotations in response to the macroscopic deformation. The number of identical rotations has been shown to dominate the opposite ones when the pressure increases, which can be explained by a decrease of inhomogeneous moves, on the same manner than EH do. The decrease of the boson peak intensity and its shift are usually explained by a suppression of modes responsible for the boson peak. Another explanation could be based on the variation of inhomogeneous moves and EH when the pressure increases, as in the case of the soft LJ glass former.

Hence, compiling these results with the ones obtained from the vibrational spectrum, it has been shown that the boson peak frequency follows the same pressure dependence than that of transverse elastic waves with frequency associated with the EH characteristic size measured in that glass. This corroborates results from the simple two-dimensional soft LJ glass former, and should encourage to extend the methodology developed for the eigenvectors study, to the case of the silica glass. The task is much more harder computationally, as one has to consider systems with a large amount of particles in order to get rid of finite size effects \cite{Leonforte05}. It should be also interesting to have a more precise look at the involved structural changes when the pressure increases, and how they could be related with elastic heterogeneities.

\end{document}